\documentclass[conference]{IEEEtran}
\usepackage{cite}
\usepackage{amsmath,amssymb,amsfonts}
\usepackage{algorithmic}
\usepackage[ruled, linesnumbered]{algorithm2e}
\usepackage{bm}
\usepackage{pgffor}
\usepackage{graphicx}
\usepackage{textcomp}
\usepackage{xcolor}
\usepackage{subcaption}
\def\BibTeX{{\rm B\kern-.05em{\sc i\kern-.025em b}\kern-.08em
    T\kern-.1667em\lower.7ex\hbox{E}\kern-.125emX}}
\begin{document}
\title{Distributed Agent-Based Collaborative Learning in Cross-Individual Wearable Sensor-Based Human Activity Recognition   \\
}

\author{\IEEEauthorblockN{Ahmad Esmaeili}
\IEEEauthorblockA{\textit{Computer and Information Technology} \\
\textit{Purdue University}\\
West Lafayette, IN, USA \\
aesmaei@purdue.edu}
\and
\IEEEauthorblockN{Zahra Ghorrati}
\IEEEauthorblockA{\textit{Computer and Information Technology} \\
\textit{Purdue University}\\
West Lafayette, IN, USA \\
zghorrat@purdue.edu}
\and
\IEEEauthorblockN{Eric T. Matson}
\IEEEauthorblockA{\textit{Computer and Information Technology} \\
\textit{Purdue University}\\
West Lafayette, IN, USA \\
ematson@purdue.edu}
\and
% \and
% \and
}

\maketitle

\begin{abstract}
The rapid growth of wearable sensor technologies holds substantial promise for the field of personalized and context-aware Human Activity Recognition. Given the inherently decentralized nature of data sources within this domain, the utilization of multi-agent systems with their inherent decentralization capabilities presents an opportunity to facilitate the development of scalable, adaptable, and privacy-conscious methodologies. This paper introduces a collaborative distributed learning approach rooted in multi-agent principles, wherein individual users of sensor-equipped devices function as agents within a distributed network, collectively contributing to the comprehensive process of learning and classifying human activities. In this proposed methodology, not only is the privacy of activity monitoring data upheld for each individual, eliminating the need for an external server to oversee the learning process, but the system also exhibits the potential to surmount the limitations of conventional centralized models and adapt to the unique attributes of each user. The proposed approach has been empirically tested on two publicly accessible human activity recognition datasets, specifically PAMAP2 and HARTH, across varying settings. The provided empirical results conclusively highlight the efficacy of inter-individual collaborative learning when contrasted with centralized configurations, both in terms of local and global generalization.
\end{abstract}


\section{Introduction}

Human Activity Recognition (HAR) is a field focused to inferring human activities from raw time-series signals, involving the identification and interpretation of human actions and behaviors through the analysis of data obtained from body-worn and/or ambient sensors, specifically designed to monitor the physical movements and gestures of individuals. The applications of HAR span a broad spectrum, with notable utilization in Ambient Assisted Living (AAL), well-being management, medical diagnosis, and particularly in elderly care \cite{snoun2023deep}. In the context of elderly care, for instance, HAR primarily serves the purpose of identifying falls and monitoring Activities of Daily Living (ADLs) among the elderly population, with a focus on delivering long-term care, promoting well-being, and preserving their quality of life, especially for those living independently.

The efficacy of everyday HAR applications relies on seamless integration and utilization of wearable sensors, spanning from simple devices, such as pedometers and accelerometers, to more computationally capable ones like smartwatches. Traditionally, wearable sensors have served the purpose of collecting monitoring data and gaining insights about individuals. However, due to significant advancements in their computational capabilities in recent years, coupled with the integration of machine learning techniques as in practices like TinyML \cite{warden2019tinyml,soro2021tinyml}, they have evolved into potent instruments for real-time activity recognition and health assessment. Capitalizing on the capabilities of these advanced devices, this paper introduces a collaborative agent-based mechanism that allows distributed, on-device learning while prioritizing privacy.

The complexity of HAR tasks is closely tied to various notable challenges, including activity variability stemming from diverse human movements, data quality and noise from sensor accuracy, scalability involving device and user numbers, privacy and security for personal data, generalization to new individuals, multimodal data fusion, and adaptability to evolving user preferences and activities \cite{chen2021deep}.

Distributed and decentralized methods in HAR offer the potential for mitigating several of these challenges. For instance, through the distribution of the learning process across networked devices and enabling on-device processing, the need for centralized data storage is reduced and most privacy concerns are alleviated. Additionally, decentralization offers the opportunity to enhance scalability through sharing the computational burden among individual devices and adaptability by allowing models to evolve in real-time and better generalize across varying individuals and settings.

Multi-Agent Systems (MAS) are a class of distributed computational systems where multiple autonomous agents independently perceive their environment and engage in interactions towards individual or collective goals \cite{wooldridge2009introduction}. Unlike traditional parallel methods, which often require a central coordinator, MAS rely on distributed and collaborative approaches, enabling agents to work together while maintaining individual decision-making capabilities. MASs' inherent decentralization offers significant advantages for the design of distributed and collaborative systems, where data privacy, adaptability, and scalability are crucial considerations. Our proposed method seeks to achieve decentralization by treating each individual as an autonomous agent, each with its dedicated machine learning model and a private, locally stored dataset. In this approach, every individual agent independently trains its personal model, learning from its unique experiences. These agents then exchange insights with their peers through a presumed communication channel connecting them, contributing to the collective intelligence of the network.

The structure of the paper is outlined as follows: In Section \ref{sec:related}, we provide an overview of noteworthy human activity recognition approaches that are relevant to the objectives of this paper. Section \ref{sec:model} delves into the details of the proposed agent-based learning method. Next, in Section \ref{sec:experiments}, we explore the characteristics and performance of the model we've put forward. To bring our discussion to a close, Section \ref{sec:conclusion} summarizes the key findings of this paper and offers insights into potential directions for future research.

\section{Related Work}\label{sec:related}
Traditionally, HAR has largely adhered to centralized paradigms, characterized by the centralized collection, pre-processing, and storage of data for training machine learning models. There are multiple recent surveys, such as \cite{zhang2022deep,chen2021deep,
kong2022human} to name a few, that discuss the notable deep learning-based advancements in this domain and underscore the evolution of the field. Centralized approaches, while effective in many respects, face limitations in addressing privacy concerns and adapting to the dynamism of real-world scenarios. Such limitations accompanied by the emergence of distributed machine learning frameworks and the ubiquity of computationally capable IoT and wearable devices have resulted in a transformative shift towards distributed methods in recent years.

The Federated Learning (FL) framework, originally introduced in \cite{mcmahan2017communication}, has served as the driving force behind a number of distributed contributions to HAR in the literature. Within the FL paradigm, a shared model is trained across a network of decentralized edge devices,such as smartphones or IoT devices, without the necessity of exchanging raw data. Instead, these devices compute model updates locally and send only these updates to a central server for aggregation and model improvement while preserving data privacy. The research conducted in \cite{sozinov2018human,bettini2021personalized,xiao2021federated,ouyang2022clusterfl,zhou20222d,shen2022federated,cheng2023protohar} are among the noteworthy research endeavors that have investigated the use of FL in HAR. While all these FL-based contributions share the common principle of utilizing local data on devices, they exhibit diversity in their specific focus, ranging from intra-device model design to comprehensive system-scale analyses. 

There are some other research contributions that exploit collaborative approaches not for building and training models but for in the enhancement of data. For instance, in \cite{lyu2017privacy}, the authors have proposed a scheme called RG-RP that uses a nonlinear function and a random projection matrix to perturb the data of participants before sending it to a cloud service, where their collaborative LSTM-CNN model is trained. In the Collaborative Self-Supervised Learning (ColloSSL) technique presented in \cite{jain2022collossl} leverages unlabeled data from multiple devices worn by a user to learn high-quality features for HAR. 

Multi-agent systems have also been previously explored as a viable approach to address key HAR, with a particular focus on collaborative recognition and data generation. The works reported in \cite{mocanu2011multi,aiello2011agent,fortino2015data, jarraya2020dcr,li2022multi} are examples of the agent-based approaches. 

In \cite{mocanu2011multi}, a high-level prototype multi-agent-based architecture is proposed for human activity recognition, incorporating domain ontology and a set of six agents responsible for tasks like individual localization and data collection from diverse sources. 

The work in \cite{aiello2011agent} presents an agent-based signal processing approach for human activity monitoring, centered on wireless body sensor networks. In this model, agents represent sensor nodes, engaging in data collection, preprocessing, and transmission to a central coordinator for subsequent analysis. 

In \cite{fortino2015data}, a platform called Cloud-assisted Agent-based Smart Home environment (CASE) is introduced. Their three-layer client server architecture  utilizes a three-layer client-server architecture in which agents are responsible for feature extraction, while the cloud takes charge of discovery tasks and the creation of classifier agents.

The DCR method, as proposed in \cite{jarraya2020dcr}, usesa triad of agents with distinct roles including data collection, segmentation and feature extraction, and activity identification. Within the DCR framework, the collaboration is established throughthe agents' ability to request one another's assistance in recognizing observed activities, with results reported based on the trust levels they mutually hold. 

The work conducted in \cite{li2022multi}, the authors have introduced a multi-agent attention-based deep learning algorithm named the Multiagent Transformer Network (MATN) to address multimodal human activity recognition. In this framework, agents represent distinct information sources, and activity classes are predicted using a decentralized actor and centralized critic approach.

Conversely, the use of MAS in human activity data generation has been a subject of research, exemplified by the work presented in \cite{kamara2017massha,barriuso2018movicloud,zhao2019bim}. The primary objective of these studies is to address privacy concerns and data insufficiency in HAR by providing realistic synthetic data through agent-based simulation techniques.

The methodology presented in this paper distinguished itself from the aforementioned approaches through its complete decentralization and relying exclusively on wearable devices and their communication capabilities. In essence, model updates and collaborations in this approach eschew the need for a central or cloud-based server, hinging instead on the interactions between an agent representing an individual and its immediate neighbors. Additionally, diverging from other agent-based paradigms that assign distinct roles, the proposed model harnesses collective intelligence among agents to enhance the generalization capability of the entire network. The solution presented in this paper is not bounded to any specific machine learning model to be used by the agents, and it not only aligns intuitively with real-world environments but also exhibits more flexibility, scaling adeptly to address the intricacies of complex scenarios. The empirical validation in two multi-subject HAR datasets of varying sizes demonstrates that the agents need not possess similar datasets, whether in terms of size and distribution as it properly accommodates the non-identically distributed (None-IID) and unbalanced nature of the time series data collected by wearable sensors through inter-agent collaboration and model updates.

\section{Proposed Model}\label{sec:model}
\begin{figure}
    \centering
    \includegraphics[width=.9\linewidth]{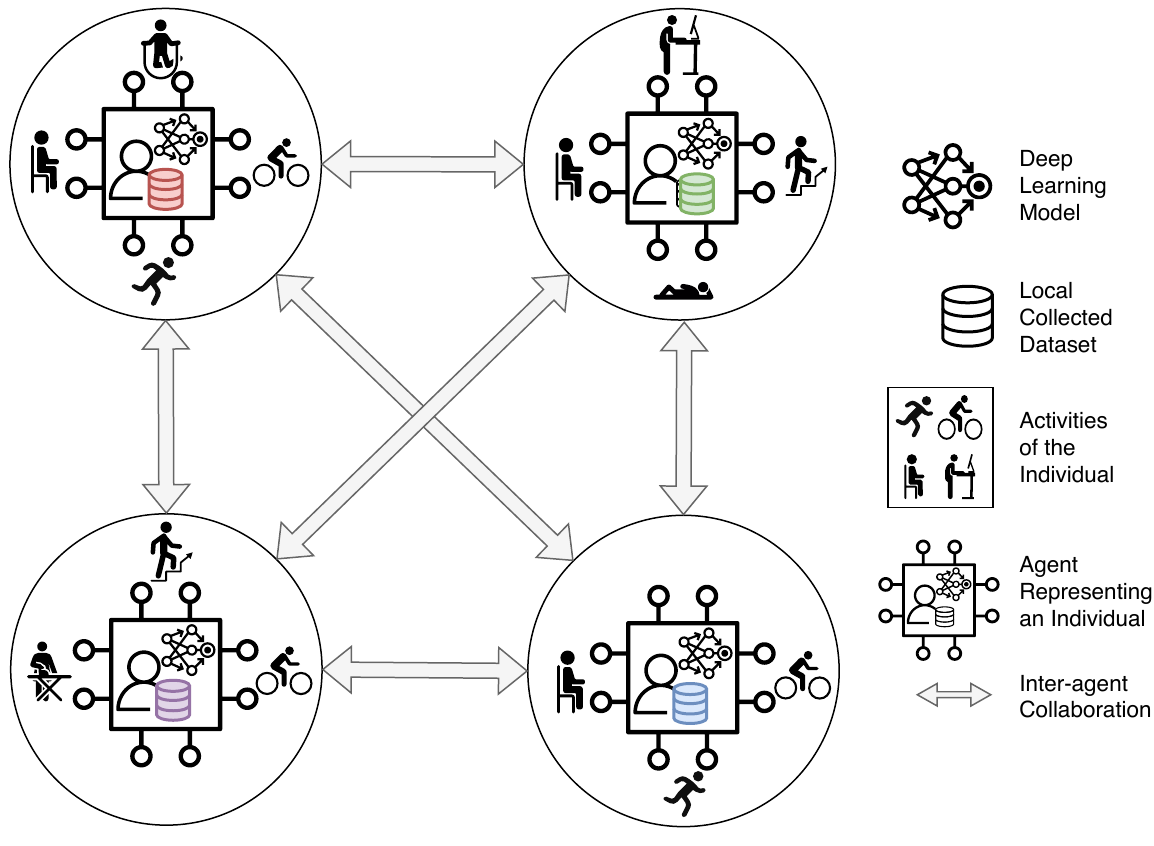}
    \caption{High-level view of a four-agent MAS for collaborative HAR in the proposed model.}
    \label{fig:harmas}
\end{figure}

This paper leverages a group of agents, each equipped with a deep learning model, to classify activities based on the multivariate sensory data collected from its corresponding individual. Figure \ref{fig:harmas} provides a high-level view of a four-agent MAS utilized for collaborative HAR. In this model, each agent functions as a processing unit representing an individual engaged in a range of physical activities. These processing units can take the form of either wearable devices with enhanced computational capabilities or dedicated Inertial Measurement Units (IMUs) connected to various sensors. As depicted, the activities of each individual may vary in both quantity and variability, resulting in distinct local data. Having that said, the deep learning model employed by each agent for recognizing the activities of its corresponding individual maintains a uniform architecture across all peers and operates within the same feature space.

Formally, let $\mathcal{A}=\{a_1, a_2, \dots, a_n\}$ represent the set of agents, each corresponding to one of the $n$ individuals. We denote the MAS network for activity recognition as $\mathcal{N}\left<\mathcal{A}, \mathcal{G}\right>$, where $\mathcal{G}$ signifies the interaction graph that delineates the connections between these agents. In our approach, each agent $a_i$ is characterized by deep learning model $\mathcal{M}_i\left<\bm{\theta}_i,\bm{\lambda}_i\right>:\mathbb{R}^{|\mathcal{S}_i|}\rightarrow\mathbb{Z}_{+}$ and dataset $\mathcal{D}_i=\{(\bm{x}_i^{(t)}, {y}_i^{(t)})\}_t$, where $\bm{\theta}_i$ and $\bm{\lambda}_i$ represent the agent's learning model parameters and hyperparameters, and $\bm{x}_i^{(t)}$  denotes the value vector collected from $|\mathcal{S}_i|$ sensors worn by the agent while performing activity $y_i^{(t)}$at time step $t$. Note that, for the sake of presentation simplicity, we have assumed that activities are indexed using positive integer values. The underlying assumption of uniformity across models and feature spaces implies that:

\begin{equation}
    \forall a_i, a_j\in \mathcal{A}: \mathcal{M}_i\simeq\mathcal{M}_j \text{ and }  \mathcal{S}_i = \mathcal{S}_j
\end{equation}
where $\mathcal{M}_i\simeq\mathcal{M}_j$ signifies the structural similarity between two models $\mathcal{M}_i$ and $\mathcal{M}_j$. 

Each agent $a_i$'s personal learning process involves the minimization of the empirical risk function $\mathsf{R}_{i}$ defined based on the loss function $\mathcal{L}_i$ as follows:

\begin{equation}
     \min_{\bm{\theta}_i} \left[\mathsf{R}_{i}(\mathcal{M}_i):=\frac{1}{|\mathcal{D}_{i}|}\sum_{(\bm{x}_i^{(t)}, {y}_i^{(t)})\in \mathcal{D}_i}^{ }\mathcal{L}_{i}(\mathcal{M}_i(\bm{x}_i^{(t)});{y}_i^{(t)})\right]
 \end{equation}

Assuming that each agent independently trains its model on its locally collected sensory data, the resulting model may exhibit limited generality and will struggle to recognize activities the individual has not encountered or for which there is minimal data available. To address this challenge, we harness collaboration by sharing model parameters among neighboring agents. In essence, the proposed model seeks to approximate a near-optimal solution for the aforementioned optimization problem by employing a sequence of local model optimizations and scheduled aggregation of the models from connected peers.

Formally, we assume weighted directional connections between the agents and define the interaction weights $\mathcal{W}$ as a square matrix as follows:

\begin{equation}
    \mathcal{W}:=
    \begin{bmatrix}
        w_{1,1} & w_{1,2} & \ldots &w_{1,n}\\
        w_{2,1} & w_{2,2} & \ldots &w_{2,n}\\
        \vdots & \vdots & \ddots & \vdots\\
        w_{n,1} & w_{n,2} & \ldots &w_{n,n}\\
    \end{bmatrix}_{n\times n}
\end{equation}
where $w_{i,j}$ represents the weight that agent $a_j$ assigns to the contribution from agent $a_i$ and is defines as follows:

\begin{equation}
    w_{i,j}= \begin{cases}
        0, & \text{If } (a_j,a_i)\notin \mathcal{G}\\
        w\in\mathbb{R}_{+}, & \text{otherwise}
    \end{cases}
\end{equation}
Note that we have assumed the notation $(a_j,a_i)$ to specify a directional communication link from agent $a_j$ to agent $a_i$ within the graph $\mathcal{G}$. Practically, the agents only keep track of their immediate neighbors. With that said, using a similar notation, we employ the column vector $\bm{w_i}\in\mathbb{R}^z_{+}$ to represent the weights that agent $a_i$ assigns to its $z$ immediate contributing neighbors.

In the proposed model, each agent uses weighted averaging as the aggregation method. Formally, the aggregation function of the agent $a_i$ is denoted by $\mathsf{F}_i$ and is defines as follows:

\begin{equation}
    \mathsf{F}_i(\bm{\Theta},\bm{w_i}):=\frac{1}{\bm{w_i}^\top \bm{1}_z}(\bm{\Theta} \bm{w_i})
\end{equation}
where $\bm{1}_z$ represents an $z$-dimensional column vector of ones, and $\bm{\Theta}\in\mathbb{R}^{p\times z}$ is a matrix composed of model parameters of length $p$ from agents $z$ neighbors. That is:

\begin{equation}
    \bm{\Theta}:=
    \begin{bmatrix}
        \bm{\theta}_1& \bm{\theta}_2& \ldots& \bm{\theta}_z
    \end{bmatrix}:=
    \begin{bmatrix}
        \theta_{1,1}& \theta_{1,2}& \ldots& \theta_{1,z}\\
        \theta_{2,1}& \theta_{2,2}& \ldots& \theta_{2,z}\\
        \vdots& \vdots& \ddots& \vdots\\
        \theta_{p,1}& \theta_{p,2}& \ldots& \theta_{p,z}\\
    \end{bmatrix}_{p\times z}
\end{equation}

The weights that agents assign to their peers can be adjusted dynamically, and they may even be learned based on the interaction experiences of the agents. In this paper, we consider bidirectional communication between the agents and establish the weights based on the size of the datasets as follows:

\begin{equation}
    \mathcal{W}:=
    \begin{bmatrix}
        |\mathcal{D}_1| & |\mathcal{D}_1| & \ldots &|\mathcal{D}_1|\\
        |\mathcal{D}_2| & |\mathcal{D}_2| & \ldots &|\mathcal{D}_2|\\
        \vdots & \vdots & \ddots & \vdots\\
        |\mathcal{D}_n| & |\mathcal{D}_n| & \ldots &|\mathcal{D}_n|\\
    \end{bmatrix}_{n\times n}
\end{equation}
It's worth emphasizing that these weights are communicated during interactions, and the agents are not required to have omniscient access to all the network weights in $\mathcal{N}$ beforehand.

The specific steps that each agent follows during collaborative learning are outlined in Algorithm \ref{alg:collearn}. In line \ref{alg:collearn:should}, the function \textsc{ShouldContinue} determines when to conclude the training process. This function can be implemented using criteria such as a fixed number of training epochs or performance on the validation set as examples. The function \textsc{IsFirstIteration} in line \ref{alg:collearn:first} ensures that the agent conducts an initial training on its private dataset before initiating the collaboration process. In case that the agent is not in its first training iteration, the receiving process and the collection of the model parameters from its immediate neighbors are managed employing the function \textsc{CollectContributions} in line \ref{alg:collearn:collect}, and the aggregation result is applied to its current model. Finally, using the function \textsc{SendModel} in line \ref{alg:collearn:send}, the agent sends its model parameter values to its neighbors as part of the collaboration. This algorithm is presented in a general manner, without imposing any specific implementations. The next section of the paper provides detailed insights into the implementations utilized for conducting the experiments.

\begin{algorithm}
    \SetAlgoLined
    \DontPrintSemicolon
    \SetKw{KwPar}{in parallel}
    \SetKw{KwNBPar}{non-blocking in parallel}
    \SetKw{KwBreak}{break}
    \SetKw{KwOr}{or}
    \SetKw{KwAnd}{and}
    \SetKwProg{myproc}{Procedure}{}{end}
    \SetKwFunction{learn}{\textsc{Learn}}
    \caption{The collaborative learning procedure run by each agent $a_{i}$ in parallel.}
    \label{alg:collearn}
    \myproc{\learn{}}{
        \While{\textsc{ShouldContinue}$( )$}{\label{alg:collearn:should}
            \If{\textsc{IsFirstIteration}$( )=False$}{\label{alg:collearn:first}
                $\bm{\Theta},\bm{w_i}\gets \textsc{CollectContributions} ( )$\;\label{alg:collearn:collect}
                $\bm{\theta}_i\gets\textsf{F}_i(\bm{\Theta},\bm{w_i})$\;
            }
            $\bm{\theta}_i\gets\textsc{Train}(\mathcal{M}_i)$\;
            \For{$z \in \textsc{Neighbors}()$ \KwPar}{
                $\textsc{SendModel}(z,\theta_i)$\;\label{alg:collearn:send}
            }\label{ln:alg1-share-end}
        }
    }
\end{algorithm}

\section{Results and Discussion}\label{sec:experiments}
As mentioned earlier, our collaborative learning model for HAR is not constrained to a specific deep learning model. With that in mind, this section establishes a simplified scenario to examine the system's effectiveness and behavior under various conditions. We intentionally selected a basic deep learning model to underscore the collaboration aspects of the model. It is important to note that employing more complex learning structures within an agent's design would likely yield higher performance.

\subsection{Agent Design}
\begin{figure}
    \centering
    \includegraphics[width=.7\linewidth]{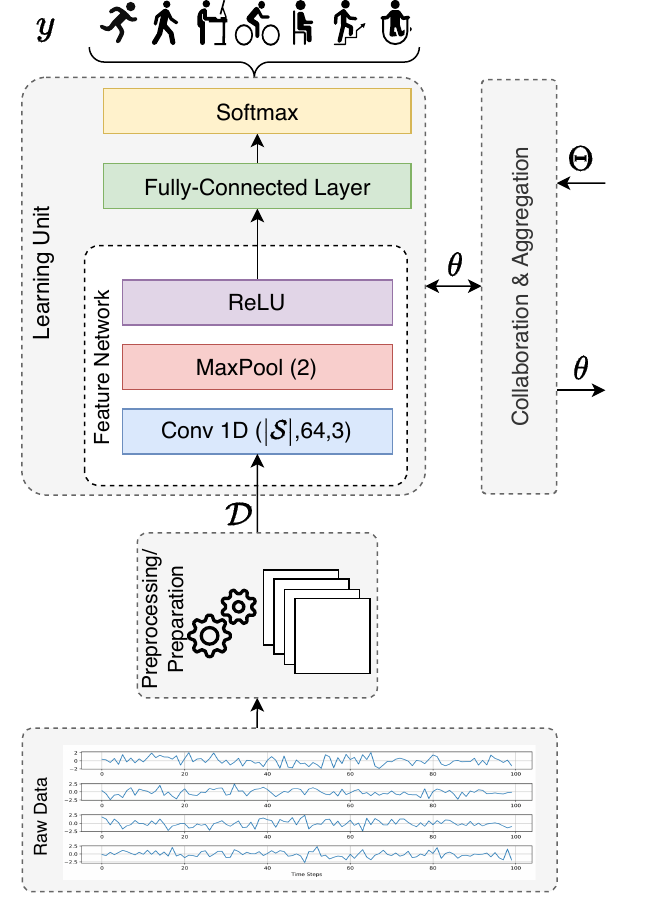}
    \caption{The building blocks of an agent's architecture.}
    \label{fig:agentmodel}
\end{figure}

As illustrated in Figure \ref{fig:agentmodel}, the fundamental components of an agent consist of three major elements: the \emph{preparation/preprocessing} unit, which performs cleansing and windowing operations on the raw multi-variate time-series data collected from the sensors, the \emph{learning} unit that holds the specification of agents deep learning model $\mathcal{M}$ and handles the learning process, and \emph{collaboration \& Aggregation} component that controls the communications with the neighbors and aggregates the model parameters as explained before.

In all our experiments, the deep learning model we utilized incorporates a convolutional feature network, consisting of a 1-D convolutional layer and a max-pooling layer, followed by a fully connected layer. The input channel of the convolutional layer corresponds to the number of features in the input data space, denoted as $|\mathcal{S}_i|$, with 64 output channels and a filter size of 3. The kernel size used for the max-pooling layer is set to 2.

\subsection{Datasets}
We have studies the behavior of the proposal model on two publicly avaiable HAR datasets explained below.
\subsubsection{PAMAP2}
Provided by \cite{misc_pamap2_physical_activity_monitoring_231}, this dataset features data collected from 9 subjects (1 female and 8 males). These subjects wore 3 Inertial Measurement Units (IMUs) on their wrists, chests, and ankles, in addition to a heart rate monitor, while performing 18 different daily physical activities, including 6 optional activities. The dataset encompasses all recorded values sampled at 100 Hz from the sensors, resulting in a raw dataset with 52 columns for each of 3,850,505 instances. In our experiments, we focused on using 36 columns, which comprise the sensory data from gyroscopes, magnetometers, and accelerometers. Additionally, the experiments excluded more complex activities such as ``playing soccer,'' ``car driving,'' ``house cleaning,'' and others, concentrating instead on the classification of remaining 12 distinct activities.

\subsubsection{HARTH}
The Human Activity Recognition Trondheim (HARTH) dataset \cite{misc_harth_779} is gathered from 22 subjects in a free-living setting wearing two 3-axial accelerometers on their right thighs and lower backs. The dataset features 6,461,328 instances sampled at a rate of 50Hz, and its 12 activities were annotated through analyzing video recordings from a chest-mounted camera frame-by-frame. In our experiments, we opted to use all 6 features from the original data and focused on 5 locomotion activities: ``walking,'' ``running,'' ``standing,'' ``sitting,'' and ``lying.''

\subsection{Experimental Settings}
The collaborative HAR system proposed in this study was developed using the \texttt{pytorch} framework, and all experiments were conducted on a computing cluster node that featured an A10 GPU, 256 GB of memory, and a 32-core CPU. All agents employ a sequence length of 100 time steps for segmenting the samples into windows within the processing unit. They utilize mini-batch training with epoch-based termination criteria in their learning units, employing a batch size of 64 windows during the training process. For the sake of optimizing the parameters of the neural network, we have used Adam method \cite{kingma2014adam} with suggested default settings $\alpha=0.001$, $\beta_1=0.9$, $\beta_2=0.999$, and $\epsilon=10^{-8}$. Collaboration between agents and their neighbors takes place at the end of each batch of training, and we have assumed a fully connected bidirectional network topology among the agents, allowing each agent to communicate with all others.

\subsection{Results}
We have assessed the effectiveness of collaborative learning in terms of both global and local generalization capabilities. Our global generalization results showcase how the trained models of the agents perform when applied to data from individuals other than those they were initially trained on. In contrast, local generalization examines how collaboration with peers impacts the generalization of the trained model when applied to the agent's private data. We have used F1-score with macro averaging as the metric to measure generalization.

\begin{figure}
    \centering
    \includegraphics[width=\linewidth]{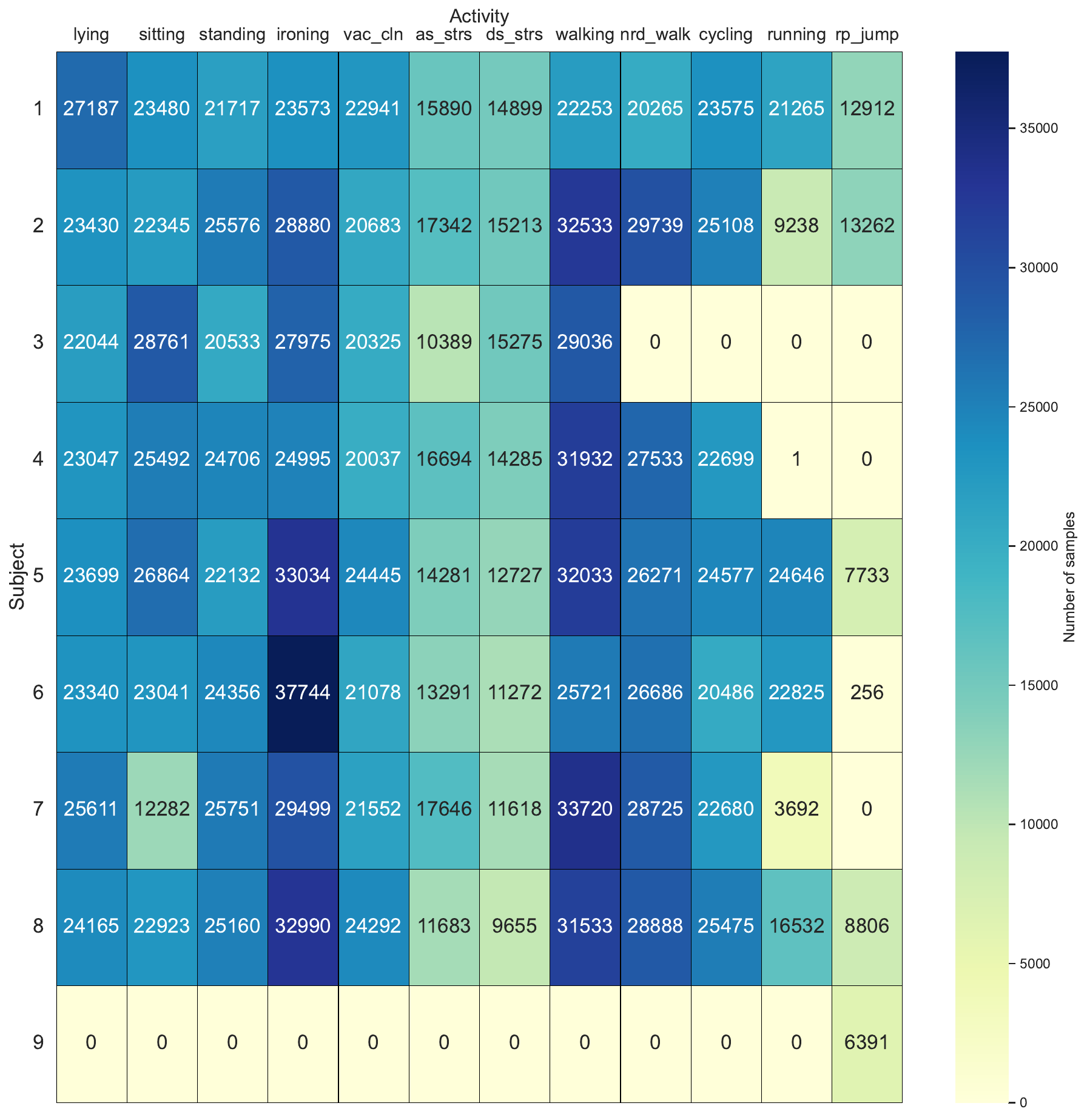}
    \caption{The data distribution for each activity within each subject in the PAMAP2 dataset used in our experiments.}
    \label{fig:pamap2_data_dist}
\end{figure}

\begin{figure}
    \centering
    \includegraphics[width=\linewidth]{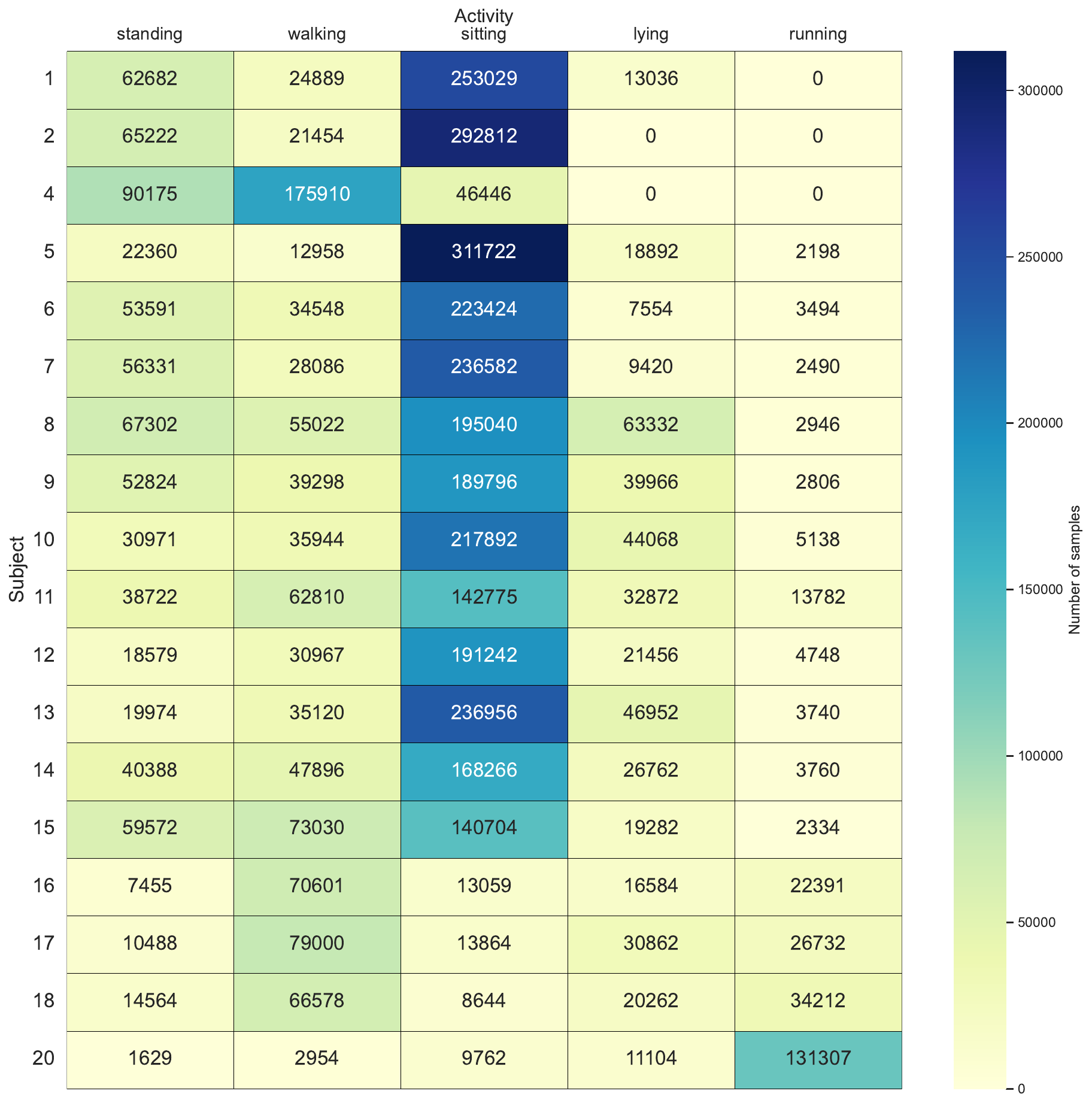}
    \caption{The data distribution for each activity within each subject in the HARTH dataset used in our experiments.}
    \label{fig:harth_data_dist}
\end{figure}

To evaluate the global generalization of the collaborative learning approach on the PAMAP2 dataset, data from subject 1 were designated as the test dataset, while data from the remaining 8 subjects was allocated for training, with each subject assigned to a separate agent. Similarly, for evaluating the global generalization of the HARTH dataset, subjects 1, 2, 4, 5, 6, 7, 8, 9, 10, 11, 12, 13, 14, 15, 17, and 20 were selected as training agents, and data from subjects 16 and 18 were designated for testing. This partitioning was carried out to ensure that the test subjects had samples for all performed activities. Figures \ref{fig:pamap2_data_dist} and \ref{fig:harth_data_dist} provide an overview of the activity data distributions for each used subject in the PAMAP2 and HARTH datasets, respectively.

To investigate the local generalization capability of the proposed learning model on both PAMAP2 and HARTH datasets, we randomly divided the data for each subject into 80\% training and 20\% testing sets. Agents representing the subjects then collaboratively trained their models using the training set and assessed local generalization on their respective test sets.

In both of the previously mentioned generalization assessments, we also conducted a performance comparison between our proposed decentralized method and a centralized approach. In the centralized approach, we followed the same training and test dataset splits as described earlier, with the only difference being that a single agent was given all the samples from the training set to construct its model in isolation.

\subsubsection{Result on PAMAP2}
\begin{figure}
    \centering
    \begin{subfigure}[b]{.8\columnwidth}
        \includegraphics[width=\linewidth]{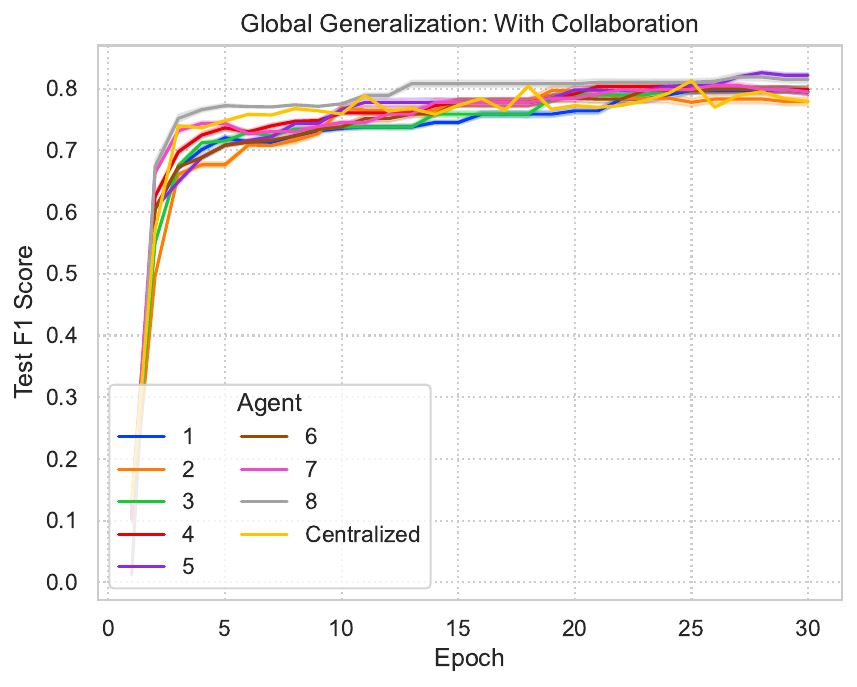}
        \vspace{-7mm}\caption{With Collaboration}
        \label{fig:pamap_global_all_w_colab}
    \end{subfigure}\vspace{5mm}
    \begin{subfigure}[b]{.8\columnwidth}
        \includegraphics[width=\linewidth]{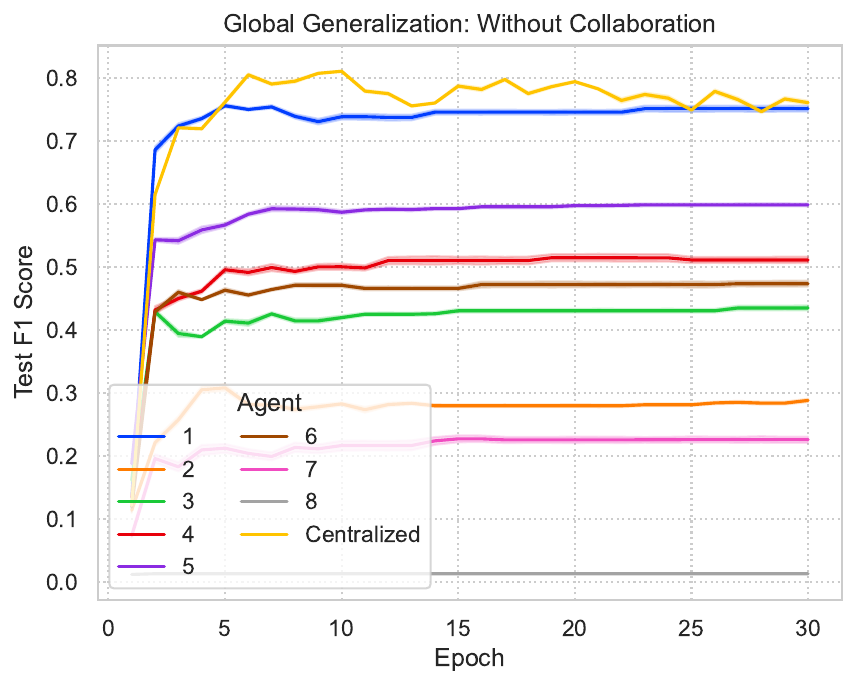}
        \vspace{-7mm}\caption{Without Collaboration}
        \label{fig:pamap_global_all_wo_colab}
    \end{subfigure}
    
    \caption{The global generalization of the proposed decentralized model on the PAMAP2 dataset.}
    \label{fig:pamap_global_all_w_wo_colab}
\end{figure}

\begin{figure}
    \centering
    \begin{subfigure}[b]{.8\columnwidth}
        \includegraphics[width=\linewidth]{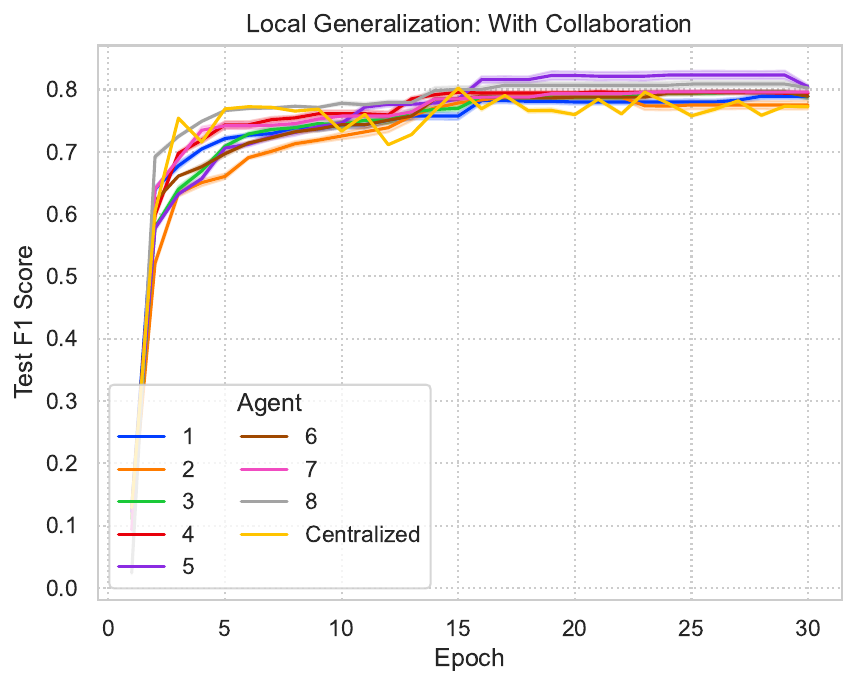}
        \vspace{-7mm}\caption{With Collaboration}
        \label{fig:pamap_loalc_all_w_colab}
    \end{subfigure}\vspace{5mm}
    \begin{subfigure}[b]{.8\columnwidth}
        \includegraphics[width=\linewidth]{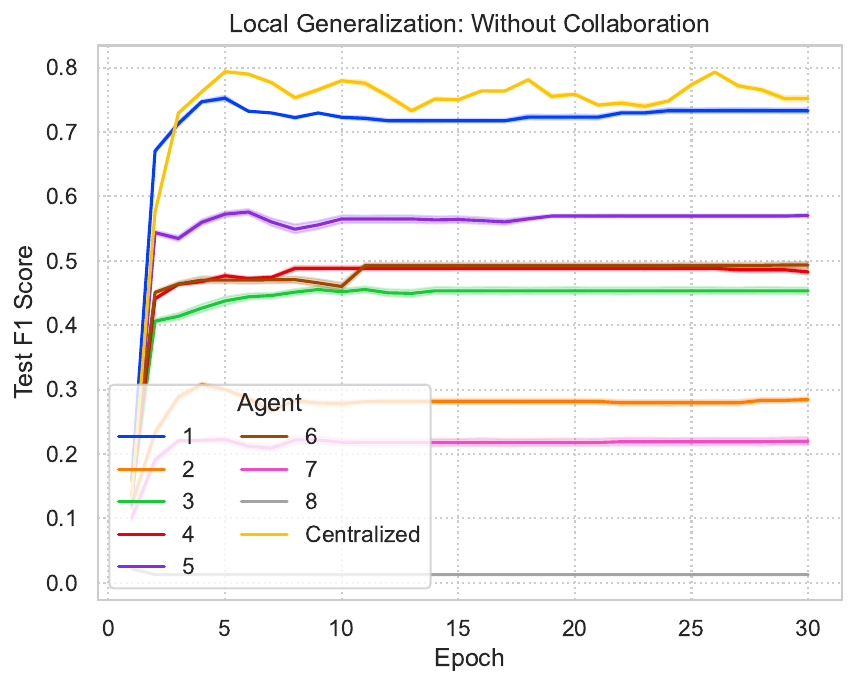}
        \vspace{-7mm}\caption{Without Collaboration}
        \label{fig:pamap_local_all_wo_colab}
    \end{subfigure}
    
    \caption{The local generalization of the proposed decentralized model on the PAMAP2 dataset.}
    \label{fig:pamap_local_all_w_wo_colab}
\end{figure}

\begin{figure}
    \centering
    \begin{subfigure}[b]{.49\columnwidth}
        \includegraphics[width=\linewidth]{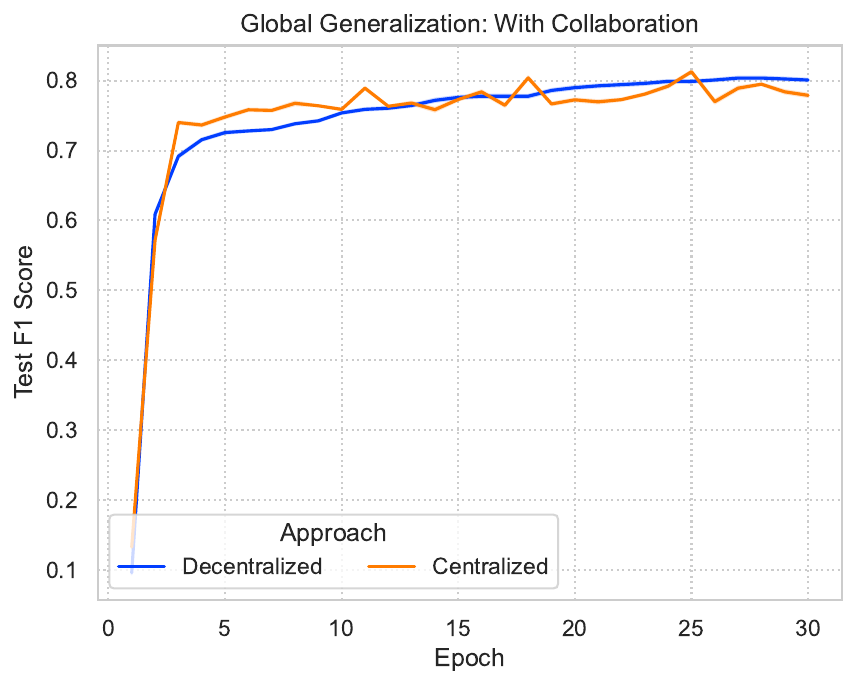}
        \caption{Global With Collaboration}
        \label{fig:pamap_global_avg_w_colab}
    \end{subfigure}
    \begin{subfigure}[b]{.49\columnwidth}
        \includegraphics[width=\linewidth]{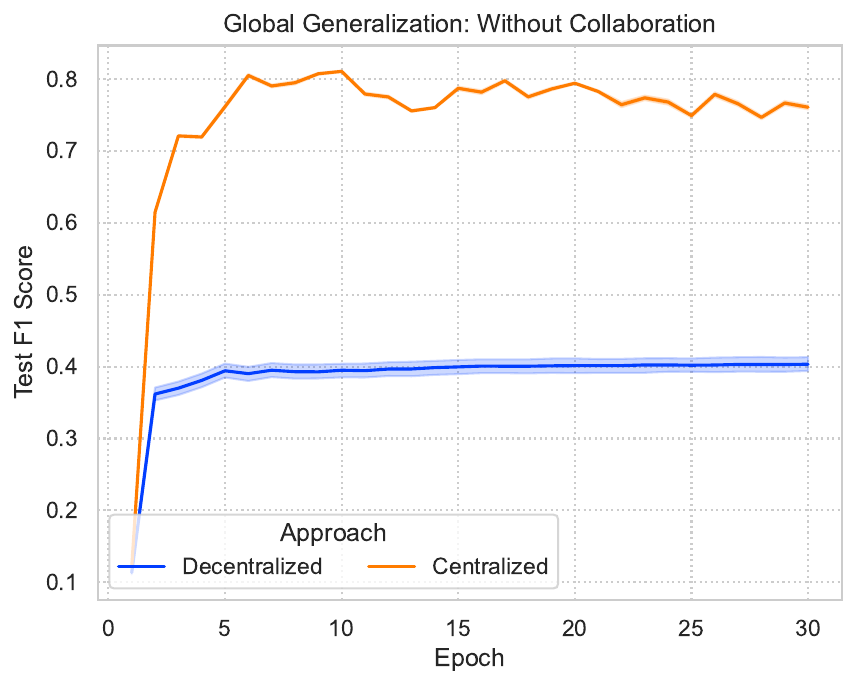}
        \caption{Global Without Collaboration}
        \label{fig:pamap_global_avg_wo_colab}
    \end{subfigure}\\
    \begin{subfigure}[b]{.49\columnwidth}
        \includegraphics[width=\linewidth]{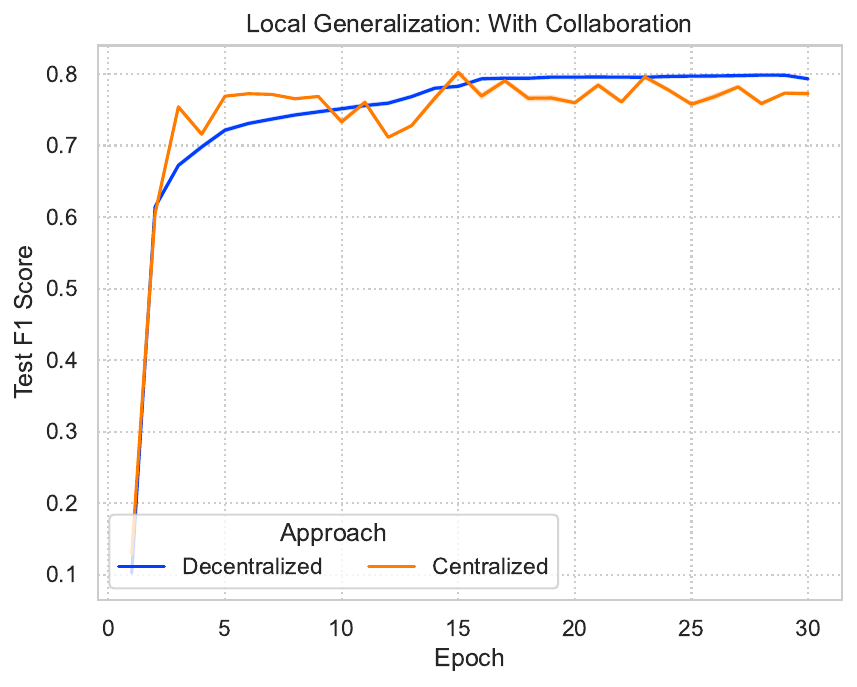}
        \caption{Local With Collaboration}
        \label{fig:pamap_local_avg_w_colab}
    \end{subfigure}
    \begin{subfigure}[b]{.49\columnwidth}
        \includegraphics[width=\linewidth]{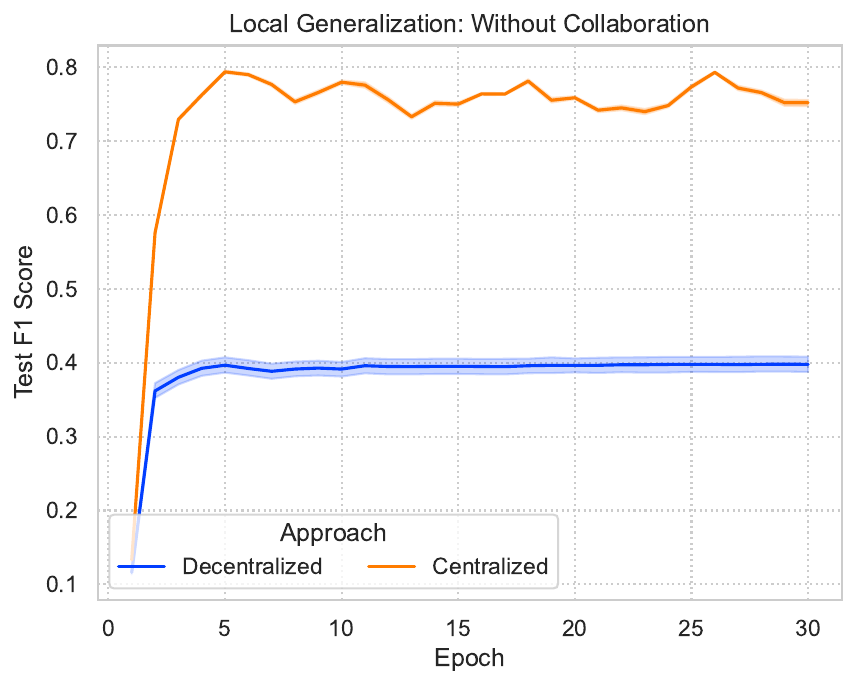}
        \caption{Local Without Collaboration}
        \label{fig:pamap_local_avg_wo_colab}
    \end{subfigure}
    
    \caption{The average performance comparison between the centralized and decentralized approaches in terms of both local and global generalization on the PAMAP2 dataset.}
    \label{fig:pamap_all_avg_w_wo_colab}
\end{figure}

The results, which evaluate the effects of collaboration on both global and local generalization measures of the agents at each training epoch, are presented in Figures \ref{fig:pamap_global_all_w_wo_colab} and \ref{fig:pamap_local_all_w_wo_colab}, respectively. Furthermore, Figure \ref{fig:pamap_all_avg_w_wo_colab} offers an average performance overview of the agents in the system under various settings, facilitating a macro-level comparison of the results. 

In addition to the evident positive impact of inter-agent collaboration on both global and local generalization, two significant observations deserve attention. The first observation pertains to the substantial effect of collaboration on the agent representing subject 9. As illustrated in Figure \ref{fig:pamap2_data_dist}, this subject had no training samples for 11 out of the 12 activities. Consequently, as observed in Figures \ref{fig:pamap_global_all_wo_colab} and \ref{fig:pamap_local_all_w_wo_colab}, the agent exhibited very weak performance, with F1-scores falling below 5\% for both local and global generalizations. However, thanks to the collaboration with other agents in the network, the agent's performance dramatically improved, even surpassing that of most other agents despite its limited training data. This improvement can be attributed to the agent having a comparable number of samples for the relatively challenging-to-classify ``rope\_jumping'' activity and the higher accumulated weight of its peers during aggregation.

The second important finding is the overall higher performance of the proposed decentralized method compared to the centralized approach in activity recognition. As shown in Figure \ref{fig:pamap_all_avg_w_wo_colab}, collaboration not only led to a nearly 100\% increase in the average F1-scores of agents for both local and global generalization but also resulted in performance surpassing that of a centralized system in the final training epochs, despite having the same training data and number of training batches.

\subsubsection{Result on HARTH} 
\begin{figure}
    \centering
    \begin{subfigure}[b]{.8\columnwidth}
        \includegraphics[width=\linewidth]{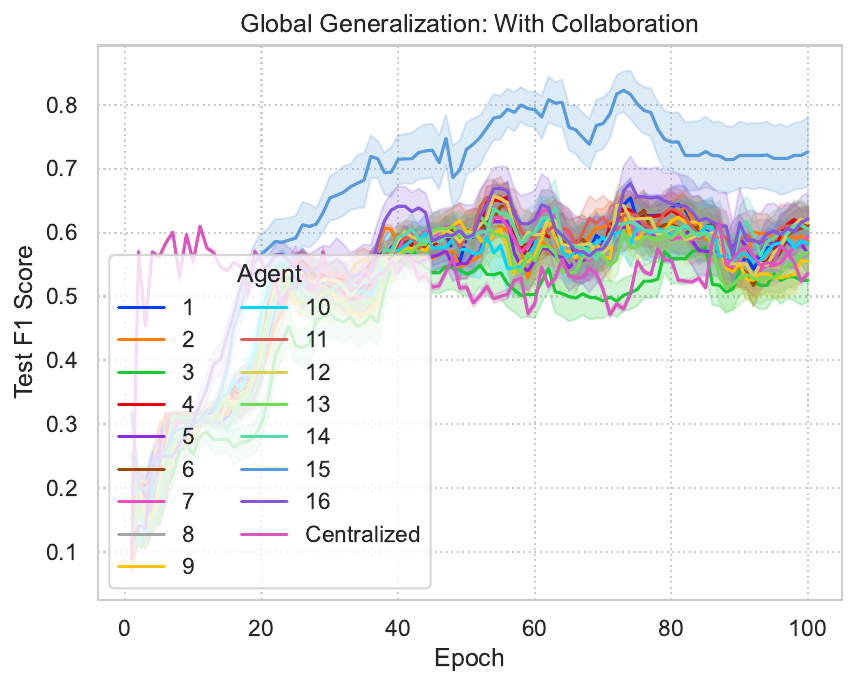}
        \vspace{-7mm}\caption{With Collaboration}
        \label{fig:harth_global_all_w_colab}
    \end{subfigure}\vspace{5mm}
    \begin{subfigure}[b]{.8\columnwidth}
        \includegraphics[width=\linewidth]{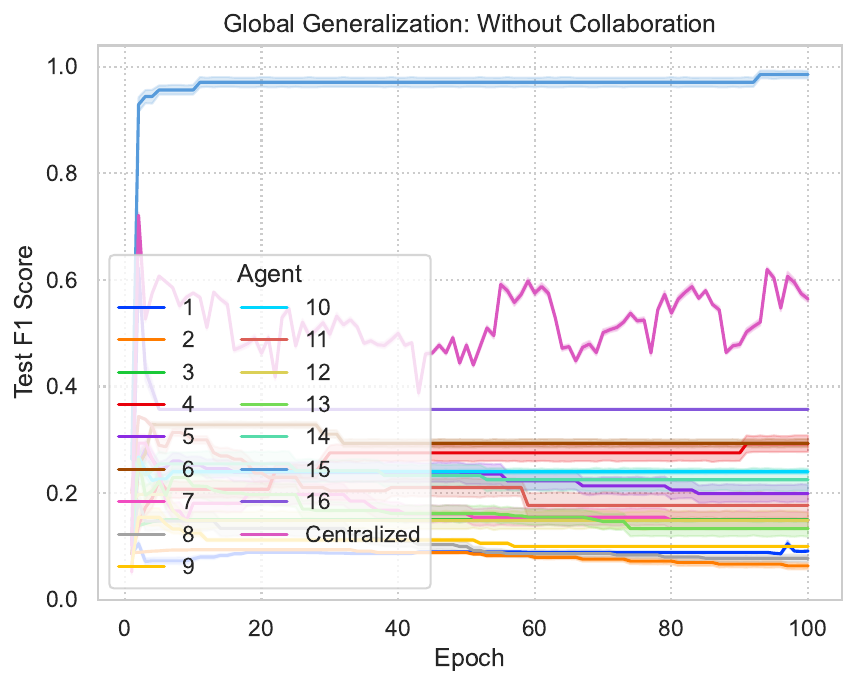}
        \vspace{-7mm}\caption{Without Collaboration}
        \label{fig:harth_global_all_wo_colab}
    \end{subfigure}
    
    \caption{The global generalization of the proposed decentralized model on the HARTH dataset.}
    \label{fig:harth_global_all_w_wo_colab}
\end{figure}

\begin{figure}
    \centering
    \begin{subfigure}[b]{.8\columnwidth}
        \includegraphics[width=\linewidth]{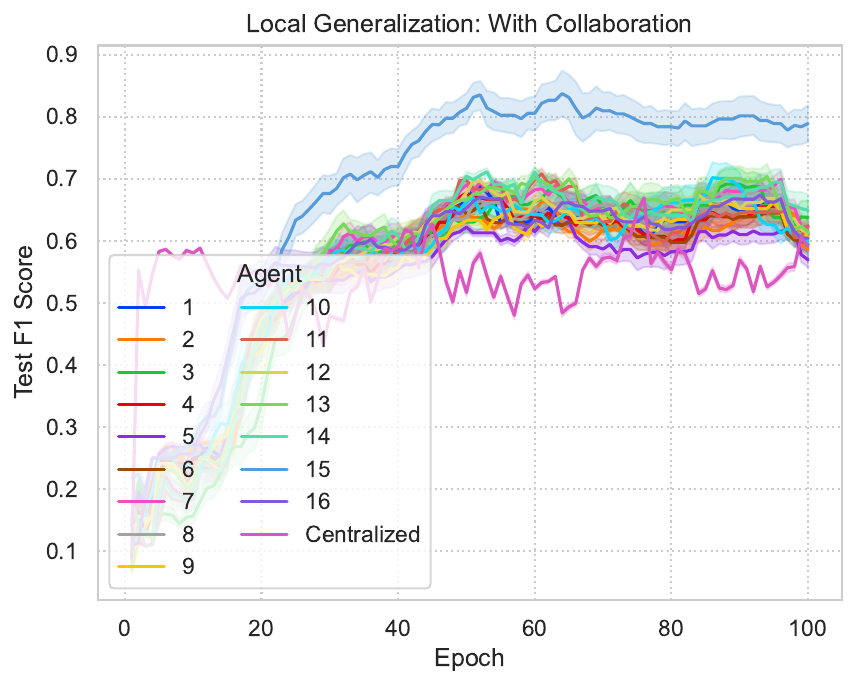}
        \vspace{-7mm}\caption{With Collaboration}
        \label{fig:harth_loalc_all_w_colab}
    \end{subfigure}\vspace{5mm}
    \begin{subfigure}[b]{.8\columnwidth}
        \includegraphics[width=\linewidth]{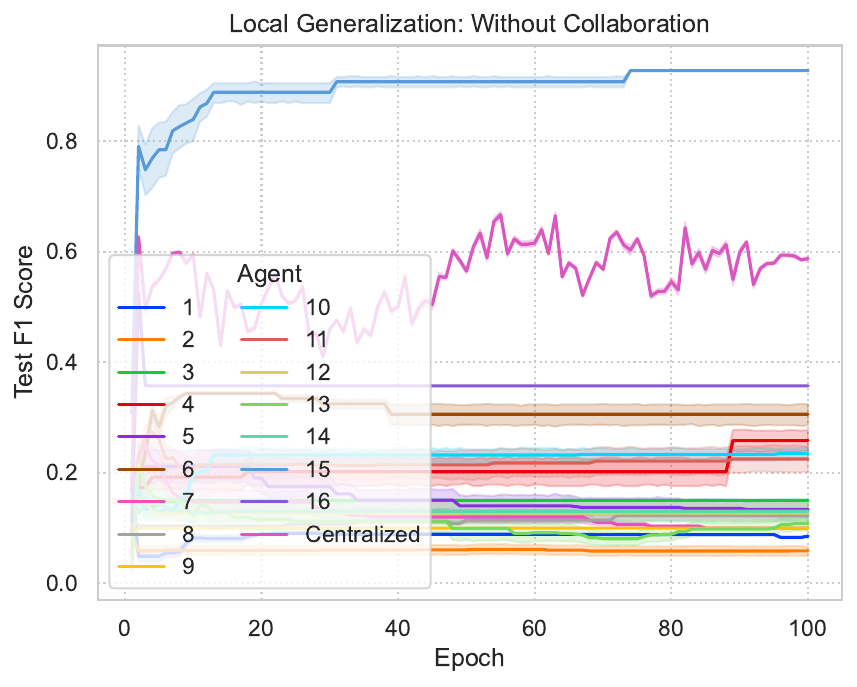}
        \vspace{-7mm}\caption{Without Collaboration}
        \label{fig:harth_local_all_wo_colab}
    \end{subfigure}
    
    \caption{The local generalization of the proposed decentralized model on the HARTH dataset.}
    \label{fig:harth_local_all_w_wo_colab}
\end{figure}

\begin{figure}
    \centering
    \begin{subfigure}[b]{.49\columnwidth}
        \includegraphics[width=\linewidth]{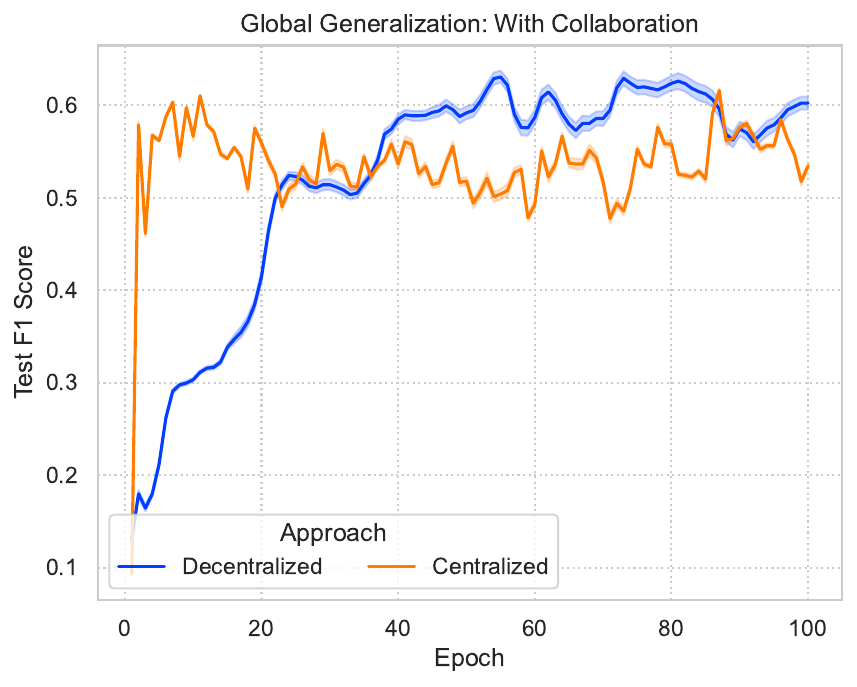}
        \caption{Global With Collaboration}
        \label{fig:harth_global_avg_w_colab}
    \end{subfigure}
    \begin{subfigure}[b]{.49\columnwidth}
        \includegraphics[width=\linewidth]{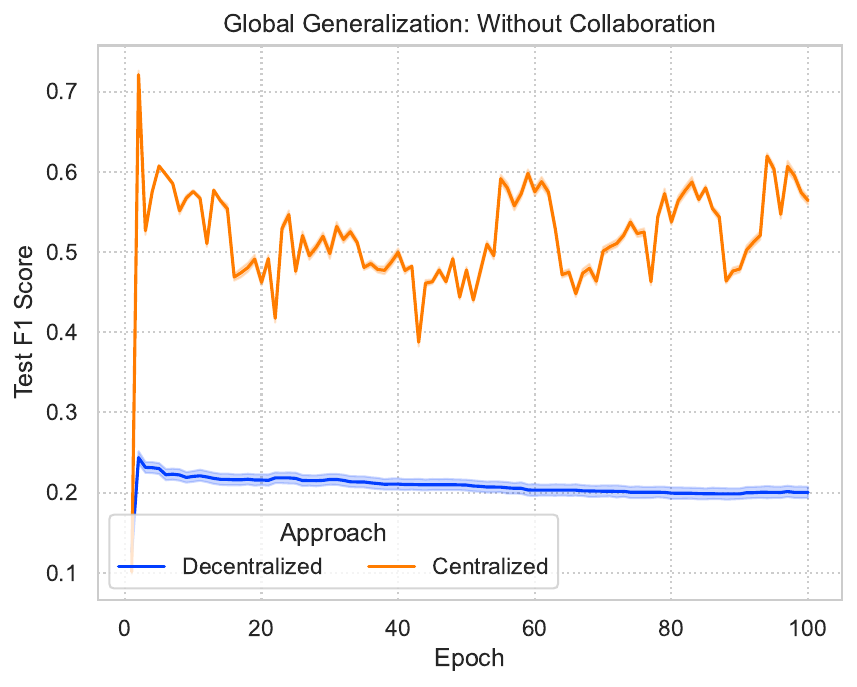}
        \caption{Global Without Collaboration}
        \label{fig:harth_global_avg_wo_colab}
    \end{subfigure}\\
    \begin{subfigure}[b]{.49\columnwidth}
        \includegraphics[width=\linewidth]{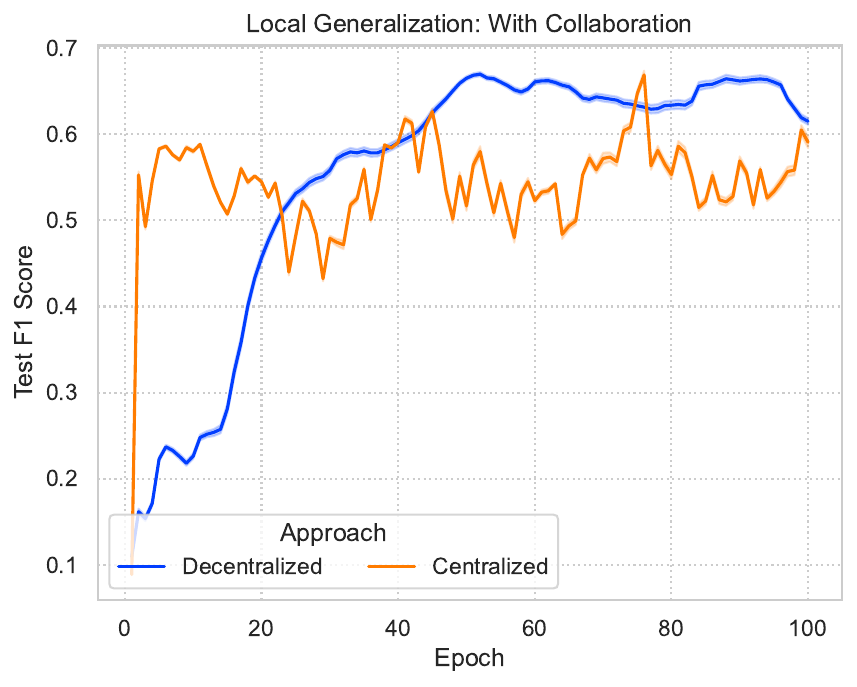}
        \caption{Local With Collaboration}
        \label{fig:harth_local_avg_w_colab}
    \end{subfigure}
    \begin{subfigure}[b]{.49\columnwidth}
        \includegraphics[width=\linewidth]{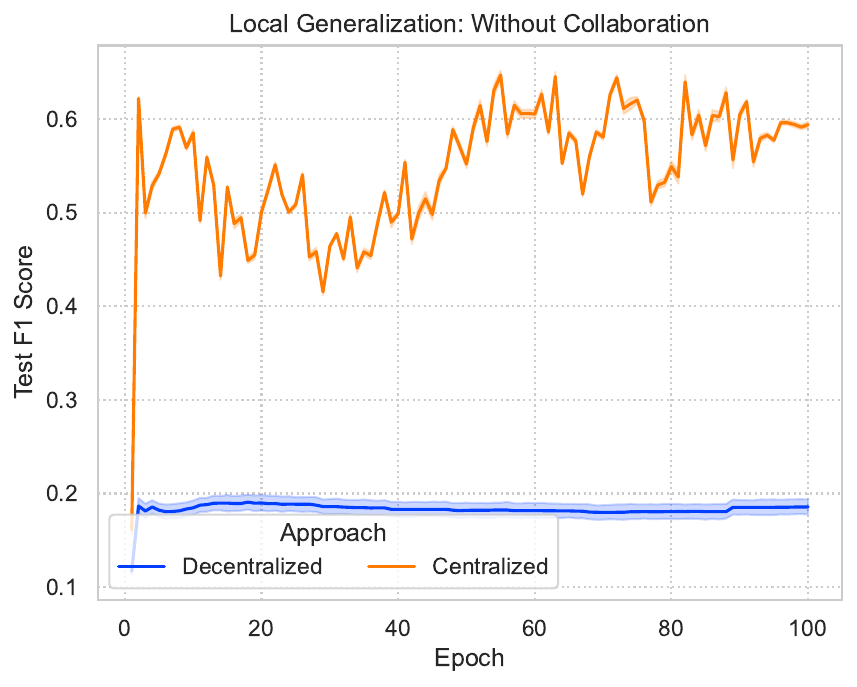}
        \caption{Local Without Collaboration}
        \label{fig:harth_local_avg_wo_colab}
    \end{subfigure}
    
    \caption{The average performance comparison between the centralized and decentralized approaches in terms of both local and global generalization on the HARTH dataset.}
    \label{fig:harth_all_avg_w_wo_colab}
\end{figure}

We conducted a similar series of experiments on the HARTH dataset and illustrated the outcomes in Figures \ref{fig:harth_global_all_w_wo_colab}, \ref{fig:harth_local_all_w_wo_colab}, and \ref{fig:harth_all_avg_w_wo_colab}. Our observations align with those made in the PAMAP2 experiments, underscoring the substantial positive impact of collaboration, especially for agents dealing with challenging training datasets. As depicted in the figures, except for agent 15, which represents subject 18 in the dataset, all agents exhibited notably improved performance compared to the centralized approach, achieving an impressive average F1-score increase of approximately 200\% when leveraging collaboration.

One intriguing aspect revealed in the results is how collaboration affected the highest-performing agent, agent 15. A comparison between results with and without collaboration shows a decline in the agent's performance, which we believe can potentially be corrected as training progresses into higher epochs and by incorporating more adaptive techniques for managing the weight assignments during the aggregation process.

It is worth highlighting that these results were achieved using a simple deep learning model within the agents, without extensive fine-tuning for the experiments. Furthermore, since the network size in the HARTH experiment is nearly twice that of the PAMAP2 experiment, it is expected that convergence and improvements in the collaborative system would proceed at a slower pace, contributing to the relatively pronounced performance oscillations observed as the training epochs advanced. 

\section{conclusion}\label{sec:conclusion}
This paper introduced a decentralized approach to Human Activity Recognition. The proposed agent-based method leverages the collaborative capabilities of individual agents, each equipped with its deep learning model, to collectively recognize a diverse range of activities from multi-variate time series data. We conducted comprehensive experiments on the PAMAP2 and HARTH datasets, comparing the performance of our approach to a centralized model. The results unequivocally demonstrate the positive impact of inter-agent collaboration, significantly improving both local and global generalization capabilities. Remarkably, the decentralized method's performance surpassed that of the centralized system, even in the presence of the same training data, highlighting its potential for real-world applications.

For future work, there are several promising directions to explore. First, we aim to investigate the scalability of our approach to larger networks with more diverse agents, potentially including non-human sensors. Additionally, we intend to delve into adaptive learning strategies, where agents dynamically adjust their inter-agent connections and weights based on their experiences. Privacy preservation is another critical avenue, as we plan to extend our work to ensure robust data security within the collaborative framework. Furthermore, we aim to test our approach on more diverse and complex datasets, enabling us to validate its efficacy in a broader range of real-world scenarios.

\bibliographystyle{ieeetr}
\bibliography{Main}

\end{document}